# Effect of Leading-Edge Tubercles on Compressor Cascade Performance

**M. C. Keerthi, M. S. Rajeshwaran, Abhijit Kushari, Ashoke De**

Department of Aerospace Engineering

Indian Institute of Technology Kanpur

Kanpur, Uttar Pradesh, India

**ABSTRACT**

Tubercles are modifications to the leading edge of an airfoil in the form of blunt wave-like serrations. Several studies on the effect of tubercles on isolated airfoils have shown a beneficial effect in the post-stall regime, as reduced drag and increased lift, leading to a delay of stall. The prospect of delaying stall is particularly attractive to designers of axial compressors in gas turbines, as this leads to designs with higher loading and therefore higher pressure rise with fewer number of stages. In the present study, experiments were performed on a cascade of airfoils with NACA 65209 profile with different tubercle geometries. The measurements were made over an exit plane using a five-hole probe to compare the cascade performance parameters. Additionally, hot-wire measurements were taken near the blade surface to understand the nature of the flow in the region close to the tubercles. Oil-flow visualization on the cascade end wall reveal the flow through the passage of blades with and without tubercles. For the cascade considered, the estimated stall angle for the best performing set of blades is found to increase up to 8.6° from that of the unmodified blade of 6.0°. Application of such structures in axial compressor blades may well lead to suppression of stall in axial compressors and extend the operating range.

**NOMENCLATURE**

$l$        blade chord

$s$        blade pitch

$\xi$        stagger angle

$i$        incidence angle

| | |
|---|---|
| $\rho$ | air density |
| $\alpha_1$ | cascade flow inlet angle |
| $\alpha_2$ | cascade flow outlet angle |
| $\alpha_1'$ | cascade blade inlet angle |
| $\alpha_2'$ | cascade blade outlet angle |
| $\alpha_m$ | mean cascade flow angle |
| $\zeta$ | total pressure loss coefficient |
| $c_1$ | inlet velocity vector |
| $c_2$ | outlet velocity vector |
| $c_{x1}$ | x-component of inlet velocity vector |
| $c_{x2}$ | x-component of outlet velocity vector |
| $c_{y1}$ | y-component of inlet velocity vector |
| $c_{y2}$ | y-component of outlet velocity vector |
| $c_m$ | mean velocity vector |
| $\Delta p_0$ | total pressure loss |
| $C_f$ | tangential force coefficient |
| $C_D$ | drag coefficient |
| $C_L$ | lift coefficient |
| PS | pressure surface of a blade |
| SS | suction surface of a blade |

## 1. INTRODUCTION

Turbomachine flows are known to be one of the most complex fluid flows. The complexity is due to various influences including that the effect of numerous internal surfaces leading to viscous shear layers and the turning of flow leading to secondary flows [1]. In addition, the flow in a compressor is subjected to an adverse pressure gradient which make the blades prone to stall and, under certain flow conditions, can potentially lead to an unfavorable phenomenon called rotating stall [2]. In its simplest form, an increased flow incidence on one blade causes separation on the suction side of that blade. Due to the reduced amount of flow around that blade, the flow diverts in a direction that increases the incidence on the neighboring blade. This process repeats over the other blades which causes the initial blades to recover from stall, leading to a "cell" of stalled blades rotating in a direction opposite to the blades. This can lead to large amplitudes of fluctuating structural loads on the blades that may lead to breakage of blades. The blockages caused by rotating stall can further lead to a catastrophic phenomenon called surge. Compressor surge is a global instability that can result in strong flow field transients leading to the destruction of an engine [3]. To avoid conditions that lead to surge, a surge margin is incorporated, where the operating mass flow rate is maintained beyond a certain fraction of the corresponding stalling mass flow rate. Since the pressure rise across the compressor is highest at the stalling mass flow rate, providing a large surge margin is considered wasteful, although necessary as the stalling point depends on several inlet flow conditions. However, delaying the onset of stall can prevent stalling when compressors operate at off-design conditions, as well as enable more aggressive designs which can provide higher pressure ratios for the same power input.

The above points highlight the consequences of blade stall and suggest that the efforts to minimize blade stall are of significant benefit [4], and this is the underlying motivation for this study. In this respect, several methods are studied and employed to delay stall and ultimately to avoid having to provide a large surge margin. Stall prevention techniques mainly come under two categories: active and passive flow control. Although active control techniques such as timed air injection [5] have the advantage of being able to function over a wide range of operating conditions, they are not popular among aviation gas turbine manufacturers due to their increased complexity. Passive control methods [6, 7] such as casing treatment, are well suited to such circumstances, as they do not involve additional energy input and are more reliable.

In the present study, a method that is moderately well-recognized for stall delay in isolated wings is employed for a cascade of compressor airfoils. The method involves the modification of the geometry of the airfoil leading edges in the shape of structures known as tubercles. In the context of fluid mechanics, tubercles are rounded protuberances periodically distributed along the leading-edge of an airfoil. These structures are known to alter the local flow field by mechanisms that are not completely understood. This local change of flow is found to substantially alter the overall performance of the airfoils, by altering the separation characteristics. The idea of flow control using leading edge tubercles originated by observing the motion of aquatic animals. The high aspect ratio flippers of the humpback whale were found to produce a high degree of maneuverability [8], and the leading edge tubercles were attributed to its sustained lift at high angles of attack [9]. Fish and Lauder [10] have reviewed a number of flow control mechanisms that enhance the locomotion of aquatic animals. A key feature of drag reduction in nature is attributed to the avoidance or delay of separation [11], since it is accompanied by a large increase in drag. A fundamental study by Owen et al. [12] on a circular cylinder showed the underlying mechanism of a wavy leading edge. Using flow visualization, they showed that the von Kármán vortex sheet was suppressed when a circular cylinder with a sinuous axis was employed. In the region downstream of the peaks the wakes were narrower. Vortex loops were formed downstream of the trough where the wakes were wider. Such a formation of streamwise vortices in the trough regions is also reported in Fish and Lauder [10].

One of the earliest studies on the hydrodynamics of the humpback whale flipper was conducted as a panel method simulation by Watts and Fish [13]. The performance of a finite NACA 63-021 airfoil with sinusoidal leading edge tubercles having an aspect ratio of 2.04 was compared with an unmodified airfoil. The results reported a 4.8% increase in lift, 10.9% reduction in induced drag, and 17.6% increase in lift-to-drag ratio at a 10° angle of attack. It was also seen that at zero angle of attack the tubercles had negligible effect on drag but at angle of attack 10° it showed an 11% increase in form drag. It was theorized that the performance improvement is due to a wavy separation line. Miklosovic et al. [14] conducted one of the earliest experimental investigations on the stall delaying characteristics of the humpback whale flipper. For this study, they employed idealized scale models of the humpback whale flipper with and without tubercle modifications at Re = 505,000-520,000. The results showed a 40% increase in the stall angle, a 6% increase in the total maximum lift coefficient, and a decrease in drag up to 32% in the post-stall regime. The lift-to-drag ratio was also found to improve for blades with tubercles except between angles 10° and 12°. This essentially increased the operating range of angle of attack for the flippers. Due to similarities in the manner in which the boundary layer is energized by the streamwise vortices, such a tubercles structure was suggested as an alternate to vortex generators. As this study involved flow over a three-dimensional flipper model, another study by Miklosovic et al. [15] was conducted with both semi-span and full span wings with leading edge tubercles, thus isolating the effect of span-wise flow component. It was found that the flow over tubercles was inherently three-dimensional for both the infinite and finite wings, as the sinusoidal leading edge created a varying sweep angle along the span. The semi-span wing showed 81% lower pre-stall lift than the full-span wing, but a 65% higher post stall lift. For the semi-span wing, the pre-stall and post-stall drag reduction was found to be 6% and 400% respectively. The generation of vortices was found to be beneficial only for the finite wing planforms in the range of the Reynolds numbers (274,000–277,000) tested. Due to the benefit being considerable largely in the post-stall regime, it was suggested that tubercles may find applications in wind turbine blades, where generation of lift at low speeds and unsteady winds is a challenge [16].

A numerical study using the detached eddy turbulence model was conducted by Pedro and Kobayashi [17] of the previously mentioned experimental studies [14, 15]. For the Reynolds number range considered (15,000–500,000), it was found that the predictions agreed even in the separated flow regime. The averaged shear-stress streaklines showed that the streamwise vortices emerging from the troughs also confine the separation growth in the span-wise direction, much like the wing fences used in aircraft wings. This mechanism was suggested in addition to the effect of streamwise vortices in mixing the momentum between the freestream and the boundary layer. Another numerical simulation [17] in the Reynolds number range 505,000–520,000 of the same experiments [14, 15] showed good predictions only in the pre-stall regions. The flow pathlines revealed that for the semi-span wing with tubercles, the post-stall regime was characterized by a flow that was attached in some regions along the span, resulting in a delayed effective separation line. The surface pressure distribution was observed for wings with and without tubercles, at two sections corresponding to the tubercle wave crest and trough for the latter. It was found that the peak static pressure was the highest along the tubercle trough section and lowest along the tubercle crest. These result in a more adverse pressure gradient along the airfoil chord behind the troughs compared to the crests, leading to different separation characteristics for the flow. Such a description of the effectiveness of tubercles is in contrast to the previously mentioned explanations that draw comparisons with vortex generators and wing fences.

A similar mechanism was also identified by van Nierop et al. [19] by means of an analytical study. They argued that the tubercles cannot act in a manner similar to vortex generators as the latter necessarily requires that the vortex generators be fully submerged in the boundary layer, whereas the tubercles characteristic dimensions are well above the boundary layer thickness. Their model was based on the fact that the varying length of the chord due to tubercles, result in a varying circulation in the span-wise direction which sheds a streamwise vortex sheet. This sheet creates an additional downwash and reduces the effective angle of attack, thus delaying the onset of stall.

The performance of airfoils with tubercles has been found to be different in the absence of a sweep back angle. A study by Stein and Murray [20] on two-dimensional airfoils with tubercles had shown a reduced performance compared to unmodified airfoils, at a Reynolds number of 250,000. A detailed experimental study on a NACA $63_4$-021 airfoil modelled with leading edge tubercles was conducted by Johari et al. [21]. On a three-dimensional flipper model, the stall angle was increased and a reduction in drag in the post-stall region was observed. The tubercles were believed to achieve this by affecting the induced drag and span-wise progression of stall. In addition, tuft flow visualization and dye visualization [22] showed the formation of vortices downstream of the tubercles troughs.

Another detailed experimental study on the effect of airfoils with leading edge tubercles was conducted by Hansen et al. [23], for different airfoil shapes and a systematic variation of the tubercle parameters. Force measurements on full-span blades with various combinations of tubercle amplitude and wavelength showed that when compared to the unmodified equivalent airfoil, tubercles are more beneficial for the NACA 65-021 airfoil than the NACA 0021 airfoil. This was attributed to fact that the momentum transfer rate of laminar boundary layer is much less than that of turbulent boundary layer so the former shows more improvement because of the enhanced momentum transfer provided by the tubercle modifications. It was also found that for both airfoil profiles, reducing the tubercle amplitude leads to a higher maximum lift coefficient and larger stall angle. In the post-stall regime, the larger amplitude tubercles performed better. Reducing the wavelength also leads to improvements in all aspects of lift performance, including maximum lift coefficient, stall angle, and post-stall characteristics. The results also suggest that tubercles act in a similar manner to the conventional vortex generators. A direct numerical solution study [24] (at Re = 800) conducted on a NACA 0020 wing identified a Kelvin-Helmholtz type of instability due to the difference in velocities downstream of the tubercle crest and troughs. The ensuing counter-rotating streamwise vortices then help in energizing the fluid to delay separation. A recent experimental study [25] on a NACA 4415 in a flow with Reynolds number 120,000 show the separation suppression due to tubercles from 2-D particle image velocimetry data. At high angle of attack, their surface oil-flow visualization show a delay in separation and periodically varying features along the span strongly suggesting the occurrence of flow compartmentalization.

In the above mentioned literature it can be seen that there are various experimental and numerical studies conducted on isolated airfoils but there are no reported studies on a cascade of airfoils which is the main incentive for this study. The present study is expected to help in determining the usefulness of leading edge tubercles on compressor and turbine blades. Some studies [26, 27] have been performed on crenulated trailing edges in compressor cascades examining their effect on wake dissipation characteristics. It was found that for optimum crenulation dimensions, the cascade resulted in faster wake dissipation, low total pressure loss and high flow turning angle. Such a strong correlation between control of wakes and the cascade performance further supports the motivation of this study.

The objective of the present experimental investigation is to quantify the aerodynamic benefits of sinusoidal tubercles in a linear compressor cascade with different tubercle configurations. From the results of studies on isolated airfoils, it is expected that the blades with tubercles will result in different characteristics due to the difference in boundary layer and separation characteristics. The performance of the blades with different tubercle dimensions is evaluated using the performance parameters relevant to compressors. An improvement in the averaged performance parameters using leading edge tubercles in a cascade have been reported previously [28] and this paper reports additional pitch-wise description of the flow to understand the underlying physics of the improvement in cascade performance.

## 2. EXPERIMENTAL SETUP

### 2.1. Airfoil and Cascade Models

The study was conducted on an open-circuit low speed wind tunnel consisting of a centrifugal blower driven by a 7.5 kW electric motor. The blower draws in air from the atmosphere and discharges to the test section through a settling chamber. Following the settling chamber, a contraction section of area ratio 16 accelerates flow to a test section of dimensions 150 mm × 150 mm. The tunnel inlet velocity was set by controlling the speed of the motor using a variable frequency drive. The average turbulence intensity at the test section was measured to be less than 2 %. The compressor cascade was designed in such a way that the incidence (*i*) and the stagger angle ($\xi$) is easily changed. The cascade was fabricated from polymethyl methacrylate sheets to facilitate flow visualization experiments. The blade profile chosen for the cascade is the NACA 65209 profile, which is a standard axial compressor blade profile. The chord of the blade is chosen as 60 mm, so as to have an aspect ratio of 2.5. This value of aspect ratio is somewhat larger than typical for turbomachine cascade studies and can be expected to produce two-dimensional flow for a greater fraction of the span without being subjected to end-wall effects. The blade and cascade parameters are summarized in Table 1.

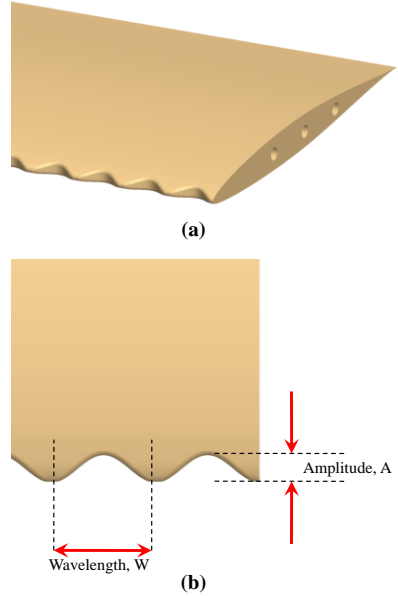

**(a)**

**(b)**

Figure 1: (a) CAD drawing showing the structure of tubercles and (b) the characteristic dimensions

Table 1: Parameters of the airfoil and the cascade

| Airfoil | | Cascade | |
|---|---|---|---|
| Chord ($l$) | 60 mm | No. of blades | 4 |
| Span | 150 mm | Blade spacing ($s$) | 30 mm |
| Aspect Ratio | 2.5 | Solidity ($l/s$) | 2 |

The tubercles structures were modeled in a CAD software by using a sinusoidal curve as the limiting leading edge. A variable radius fillet was applied to the edges, with zero radii at the crest and trough to ensure their dimensions are unaltered. The blades have a constant chord design, with constant thickness-to-chord ratio everywhere except the leading edge. That is, there are no striations along the chord, downstream of the leading edge, giving a bumpy appearance in the span-wise direction. The choice is due to the fact that modern compressor blades, which are typically subjected to large aerodynamic forces while having a low thickness, will have multiple stress concentration lines along the chord due to the bumpiness. Figure 1 (a) shows the form of the tubercles used for the present study. Figure 1(b) shows the definitions of the primary characteristic dimensions for a typical tubercle. The characteristic dimensions of the sinusoidal curve, namely its amplitude ($A$) and wavelength ($W$), were chosen so as to study the effect of amplitude and wavelength while keeping the dimensions well above the minimum tolerance of the fabrication process. The plain and modified blades were then fabricated using rapid prototyping to ensure conformity to the required geometry. At the time of fabrication, the blade was oriented in the span-wise direction. The material used for the fabrication was ABS plastic and the measured average surface roughness (measured using Mitutoyo SJ-30) was found to be 16 μm along the span, and 2 μm along the chord. Figure 2 shows a photograph of the different blades, one from every cascade studied. Figure 3 shows the schematic of the cascade turning arrangement, a photograph of which is shown in Figure 4. The details of the tubercles on the four different blades are shown in Table 2. The blades are named following the convention of Hansen et al. [23]. The apex angle of the tubercles refer to the included angle between the sides coinciding with the leading edge of the tubercle structure, equal to $2\tan^{-1}(0.5W/A)$.

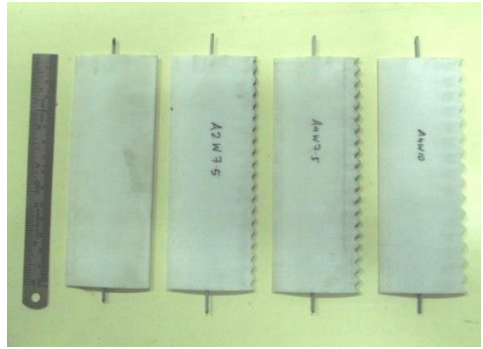

Figure 2: Photograph of the different blades

Table 2: Dimensions of the tubercles

| Cascade name | Amplitude, A | Wavelength, W | Aspect ratio, A/W | Apex angle of tubercle |
|---|---|---|---|---|
| A0W0 (unmodified) | - | - | - | - |
| A2W7.5 | 2 mm | 7.5 mm | 0.27 | 124° |
| A4W7.5 | 4 mm | 7.5 mm | 0.53 | 87° |
| A4W10 | 4 mm | 10 mm | 0.40 | 103° |

**2.2. Measurement details**

Measurements were taken at the exit plane using a five-hole pitot probe (by Aeroprobe Inc.) with tip diameter of 3.2 mm to obtain three-components of velocity, static and total pressure data over a prescribed measurement grid.

The probe is capable of measuring flows that are incident up to an angle of 60°, and a preliminary measurement showed that the flow angles remained within this limit, even downstream of the suction surface of the blades under consideration, where separation existed for certain cases. The five-hole probe was mounted on an automated two-axis traverse. An electronic pressure scanner (ESP 32 HD by Pressure Systems) of range 13.79 kPa (2 psi) was used for the measurement of all differential pressures, with the inlet freestream static pressure as the reference. The data was acquired at 256 samples per second for two seconds at each point.

The experiments were conducted at a Reynolds number of $1.3 \times 10^5$ based on the freestream inlet velocity of 35 m/s and mean chord length. The stagger angle was set at 0° for all tests. Figure 3(a) shows the schematic of the turn table with the blades, set at an arbitrary incidence. The measurements were carried out in a plane at a semi-chord distance from the trailing edges, as shown in the Figures 3(a) and 3(b). The measurement grid resolution was taken to be 5 mm in both directions, spanning an area of 100 mm (21 points along pitch-wise direction) and 125 mm (26 points along span-wise direction, with allowance for probe safety). In the present study, all results are reported for the readings taken for the central passage, i.e., between the wakes of the central two blades. The axis of the probe was aligned along the axial direction (i.e., the direction perpendicular to the line joining the leading edges of the blades) for all incidence angles. The incidence angle was adjusted using the turn table in the cascade and for $\xi = 0°$ the experiments are conducted for incidence angles of -5, -3, 1, 5, 9, 13 and 17°. To set the required incidence, the blades were turned about the cascade center, and the set angle was compared with the incoming flow direction to determine the inlet flow angle of the cascade. The incidence angle (the term used in turbomachine nomenclature) is essentially equal to the angle of attack (as defined for isolated wings) less the leading edge camber angle, as shown in Figure 3(a). The cascade was subsequently reassembled with each set of the modified blades and the experiments are repeated. For each measurement, less than 2 % of the measurement points had noise, likely originating from a data acquisition fault involving the ESP scanner, and their values were replaced by averaging the properties of the two neighboring points along the span-wise direction.

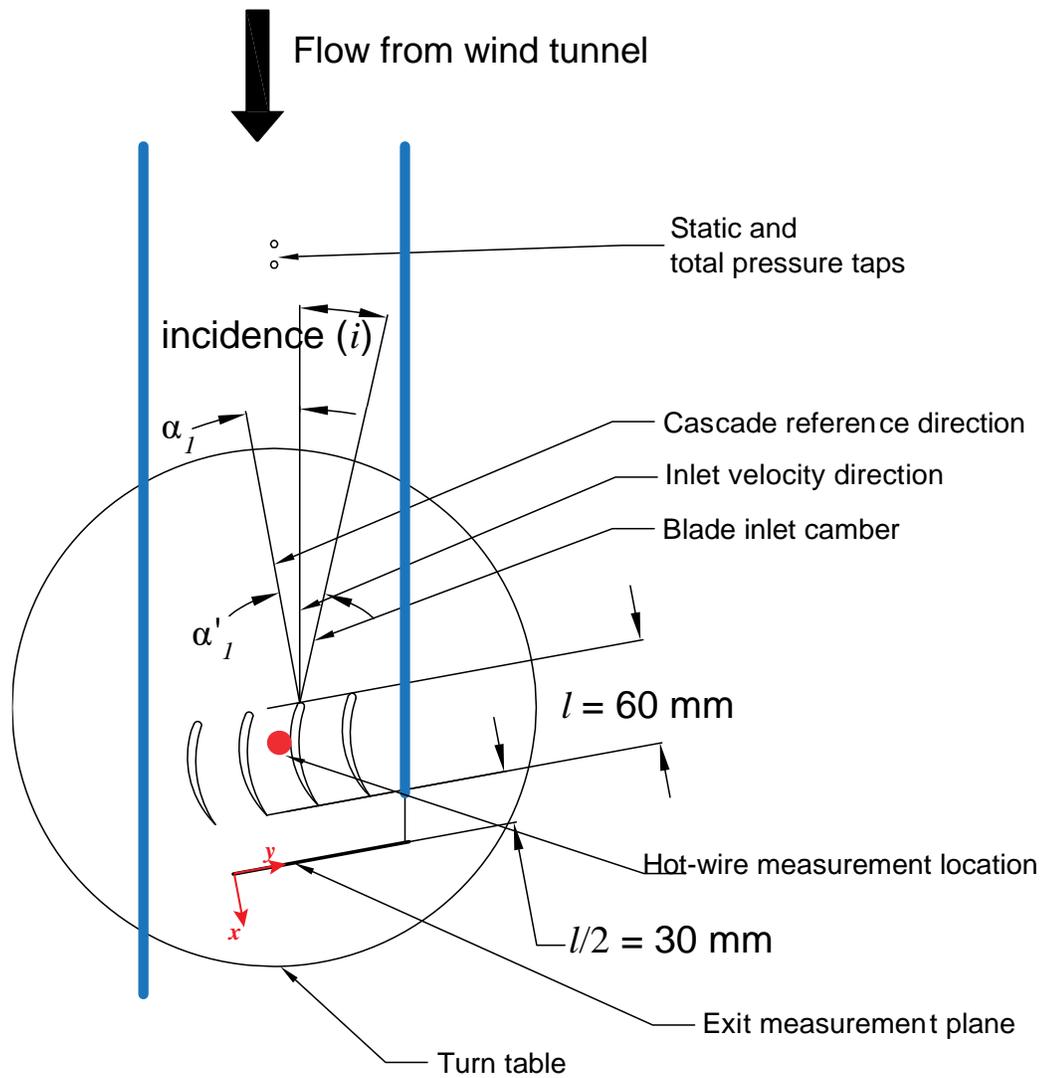

(a)

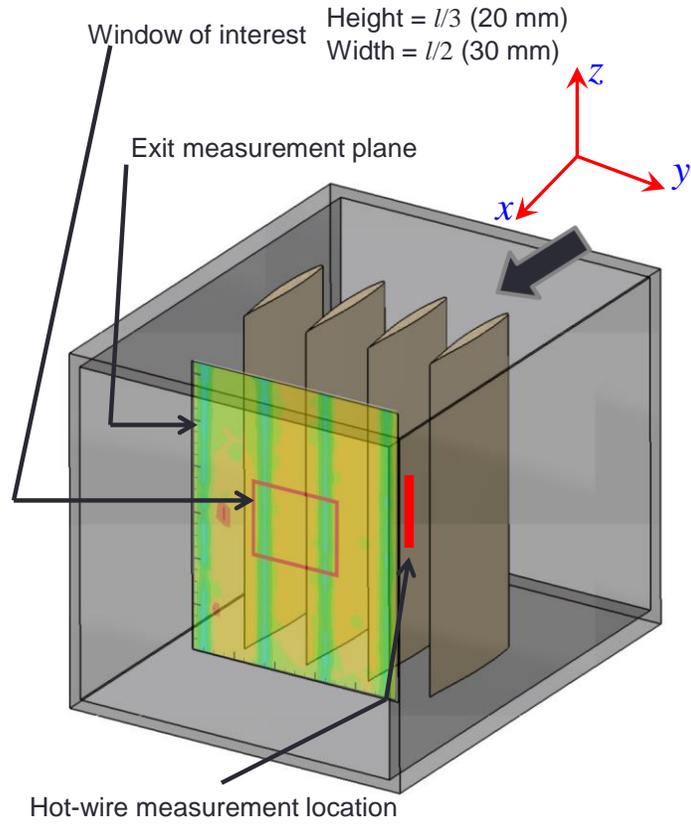

(b)

Figure 3: Schematic of the cascade arrangement showing the measurement locations for five-hole and and hot-wire probes (a) in the plan view and (b) isometric view

A single-wire miniature hot-wire probe (Dantec Dynamics probe 55P16 with the 54T30 circuitry) was used to obtain axial velocity data at selected points, as shown in Figures 3(a) and 3(b). The wire has a diameter and length of 5 μm and 1.25 mm, respectively. By holding the wire along the span-wise direction, the small characteristic length enabled measurements with superior resolution along the span-wise direction. Measurements were made at a sampling rate of 10,000 samples per second for 15 seconds, but only the mean data is reported here.

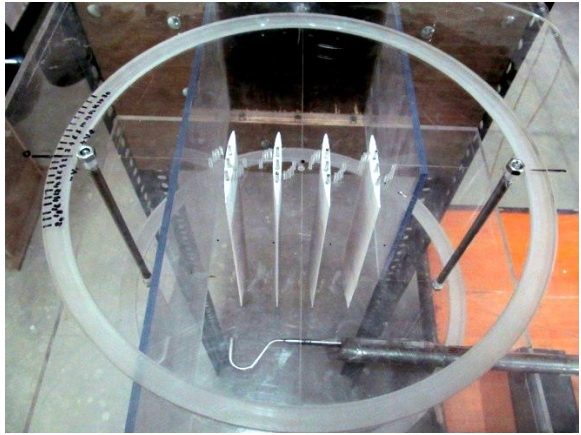

Figure 4: Photograph of the test section with the cascade set at zero incidence

In this experiment surface oil flow visualization technique was used to get qualitative data about separation. Surface flow visualization was carried out at the end walls of the cascade by employing oil film. In oil film visualization, a thin layer of oil mixed with a pigment is coated on the required surface. When the flow is established, the oil moves along the local air flow direction, leaving behind a trace of the pigment. The pattern left behind shows streaks indicative of the direction of air flow near the surface. Surface flow visualization is particularly useful to identify regions of flow separation, as the resulting pattern shows a distinct contrast from the attached flow regions.

In the present setup, a mixture of titanium dioxide powder, coconut oil and oleic acid mixed in proportions of 10:40:1 by volume [29] was used. A thin transparent plastic sheet was pasted on the lower end wall of the cascade. The solution was sprayed on the wall using a standard artist's airbrush. The blades were then fixed ensuring the oil film is undisturbed. The tunnel was switched on and the full operating speed of 35 m/s is established within a minute. The oil film stopped showing perceptible motion in about five minutes. The tunnel was kept on for a total of fifteen minutes for all the runs, after which the setup was dismantled and the transparent sheet extracted. The sheet was then placed on a black surface and the photographs were taken normal to the sheet. The end blades showed some variation between repeated runs, compared to the inner two blades, due to the side wall effect. As a result, only the flow around the central two blades is considered for analysis.

**2.3. Data reduction**

In order to estimate the lift and drag acting on a given blade from the exit plane readings, calculations on a two-dimensional airfoil cascade were employed as described in Dixon [7]. The analysis consists of applying continuity and two-dimensional momentum equations by using the following physical quantities as the inputs: the set properties, viz. inlet velocity, inlet flow angle, the blade geometry (inlet and exit blade angles and stagger), and the measured quantities, viz. exit total pressure and velocity fields. The procedure evaluates the total pressure loss coefficient ($\varsigma$), the mean flow angle ($\alpha_m$) and skin-friction coefficient ($C_f$) in the intermediate step, which are given by:

$$\zeta = \frac{\Delta p_o}{0.5 \rho c_{x1}^2} \qquad (1)$$

$$\tan \alpha_m = \frac{1}{2}(\tan \alpha_1 + \tan \alpha_2) \qquad (2)$$

$$C_f = 2(\tan \alpha_1 - \tan \alpha_2) \qquad (3)$$

Using these, the outputs, overall performance parameters namely the lift and drag coefficients are calculated, which are given as:

$$C_L = \frac{s}{l} \cos \alpha_m \left( C_f - \zeta \frac{\sin 2\alpha_m}{2} \right) \qquad (4)$$

$$C_D = \zeta \frac{s}{l} \cos^3 \alpha_m \qquad (5)$$

As the analysis uses a one-dimensional set of equations, the present measurements which are carried over a two-dimensional spatial field need to be suitably reduced. In order to get a representative value required for the above analysis, the quantities were therefore averaged along both directions. Firstly, from the raw planar field data a span-wise averaging was done, excluding a point each from the end-walls. This step essentially evaluates the column-wise average of a matrix to produce a vector. The distribution of the raw span-averaged total pressure obtained this way for a typical case (unmodified blades at incidence of 13°) is shown in Figure 5. Secondly, for the pitch-wise averaging, a range of values was extracted along the pitch, directly downstream of the second passage (the middle of the three passages), so as to effectively capture one wake. Although the passage is not identical to the geometric passage encompassed by the central blades, this choice ensures that only one wake is considered for the control volume analysis. The span-wise gradients were found to be minimal, and the variation of the span-averaged total pressure effectively captures the blade-to-blade flow properties. In Figure 5, the local minima points (at about 40, 55 and 90 mm pitch-wise distance) correspond to the wakes of the three blades, and the wake of the first blade (leftmost blade) merges with the large separation zone due to the sudden expansion along the suction surface of the first blade and side-wall of the test section. Figure 5(a) shows that the span-averaged values are reasonably periodic, and hence the central passage is expected to give a good estimate for the cascade. Periodicity was found to be better for the low incidence cases. Figure 5(b) shows the variation of standard deviation of the points along each span-wise location, whose average is reported in Figure 5(a). It can be observed that the value is relatively high in the wakes due to larger unsteadiness, but the magnitudes are acceptably low. As a large number of data points (144 points) were included in obtaining the average that is used in the analysis, it is believed that this method of data reduction produces global performance parameters that are adequately representative of a cascade with the specified parameters. It has to be noted that although the five-hole probe tip diameter is comparable to the tubercles dimensions, as the measurement plane is about 1.5 chord-lengths downstream of the leading edge, the gradients due to the fine tubercle structures can be expected to be diminished. This, combined with the fact that a large number of points are considered for averaging at the exit plane, the error in the integrated effect can be expected to be low.

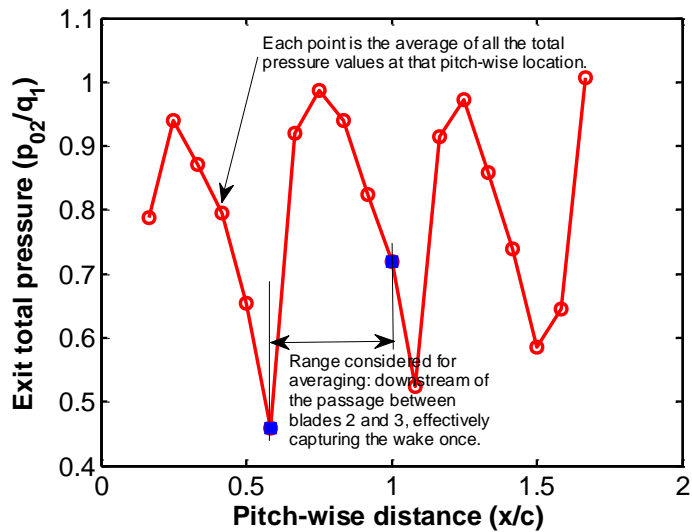

(a)

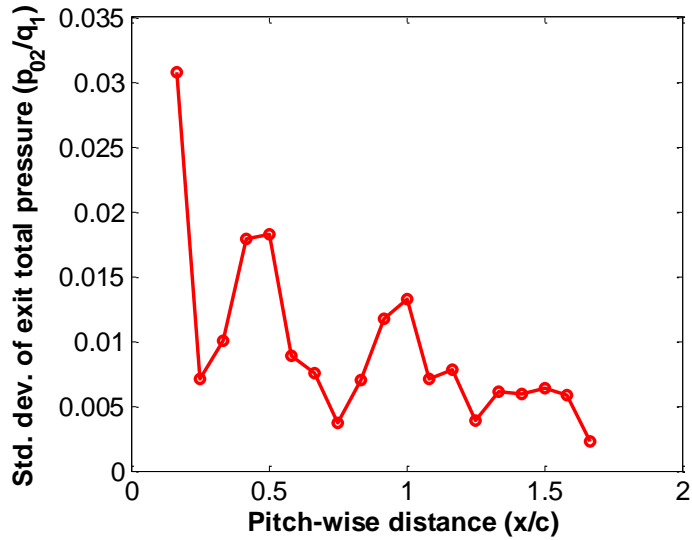

(b)

Figure 5: (a) Span-averaged total pressure values along the pitch-wise direction for a typical case (A0W0, $i = 13°$), (b) Standard deviation of the exit total pressure at each pitch-wise location

In order to quantify the uncertainty in the performance parameters, a set of 31 measurements were carried out for the central blade passage for the unmodified blades at incidence -3° and the values $C_L$, $C_D$ and mean $\varsigma$ were calculated for each of these measurements. Subsequently, the standard deviations of this sample were evaluated for these parameters and the chi-squared statistic was used to calculate the corresponding population standard deviations. These were then used to estimate the 95% confidence interval within which the parameters are expected to lie, as described in Moffat [30]. Table 3 lists the 95% confidence interval of the absolute uncertainties. The standard rules of propagation for uncertainty were used to estimate the uncertainty in stall angle from the uncertainty in mean $\varsigma$. The values were in the order of the uncertainty in typical cascade measurements [7, 31]. As it is not practical to repeat the above method of measuring multiple ensembles for each incidence and tubercles geometry considered, the uncertainties associated with this case is to be taken as representative of all cases. It has to be noted that estimating uncertainty from spatial averaging of points within a single ensemble of measurements will overestimate the uncertainty. This is because of the inherent non-uniformities present in the flow downstream of the cascade in the pitch-wise and, in case of tubercled blades, the span-wise directions.

Table 3: Uncertainty in performance parameters

| Parameter | Absolute uncertainty at 95% confidence |
|---|---|
| $C_L/C_D$ | 0.10 |
| Mean $\varsigma$ | 0.027 |
| Mean deflection | 0.2° |
| Stall angle | 0.2° |

## 3. RESULTS AND DISCUSSION

In the following sections, first, an assessment of the effectiveness of the tubercles is made by comparing the overall performance with each other and the unmodified blades. The first part (section 3.1) shows the performance of the cascade with unmodified blades. Following this (section 3.2), the performance parameters are relooked at by noting them against the change of a specific geometric attribute of the tubercles. The contours measured on the exit plane are considered next (section 3.3) in order to examine the details of the flow between blades. The direct measurement of the flow adjacent to the blade surface is then presented (section 3.4) to clarify some of the former inferences. Flow visualization results (section 3.5) are presented finally to demonstrate the working of tubercles.

### 3.1 Performance of the cascade with unmodified blades

Measurements were first made on the cascade with unmodified blades in order to determine its performance. This performance is considered as the baseline for the blades with chosen airfoil profile and cascade parameters. Figure 6 shows the variation of passage-averaged total pressure loss coefficient for different blade incidence angles. The trend shows the lowest total pressure loss occurring at a slightly negative incidence, which is a characteristic of cambered airfoils. As the incidence increases, the increasing adverse pressure gradient along the suction surface leads to flow separation, creating a wake that is distinguished as a region of reduced total pressure.

In contrast to the easily identifiable stall angle for isolated airfoils, for blades in a cascade, the drop in lift or the raise in total pressure loss with increasing incidence is gradual due to the influence of adjacent blades. This makes defining the stall angle imprecise and hence conventionally it defined as the incidence angle for which the value of $\varsigma$ reaches twice the minimum $\varsigma$ [7]. From the plot of $\varsigma$ against incidence, using a linear interpolation between the two measured incidence values, one can calculate the stall angle for a given cascade. By this definition, the unmodified blades showed a stall angle of 6.0° for the present cascade span, pitch and aspect ratio. For comparison, the stall angle for an isolated wing with the same airfoil profile and close to the present Reynolds number is reported to be about 11° [32].

The variation of passage-averaged flow deflection with incidence is shown in Figure 7. The flow deflection, defined as $\alpha_1 - \alpha_2$, is a measure of how effective the turning caused by the cambered blades is. Together with the total pressure loss coefficient, this value affects the other performance parameters, as can be seen from equations 1–5. For a cascade, the flow is deflected more as the blade incidence increased, up to the point where the separation is severe enough that the fluid in wake regions counteract the fluid turned in the freestream. Both Figures 6 and 7 show a slight irregularity in the incidence range 5–9°. This could be because of the incidence angles being close to the cascade stall onset incidence.

The variation in the lift-to-drag ratio with incidence is shown in Figure 8. Although the blade lift-to-drag ratio is difficult to interpret as a turbomachine performance parameter, it may be considered as a single parameter to evaluate the global performance of the cascade, as it includes a contribution from both total pressure loss and flow angles (Equations 4 and 5). The effect of the possible flow separation is seen here as a noticeable drop after an incidence of 5°.

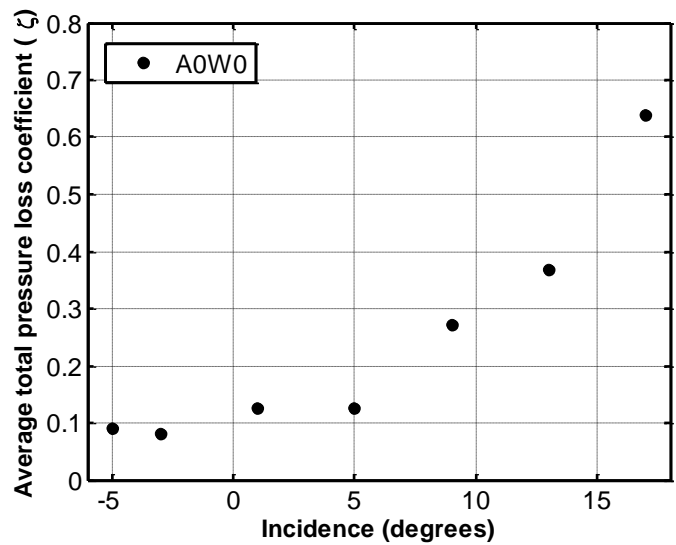

Figure 6: Variation of total pressure loss coefficient with incidence for the unmodified blades

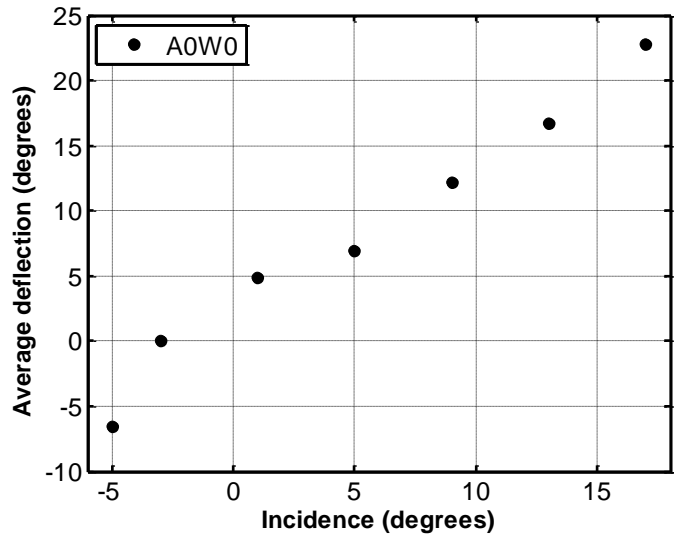

Figure 7: Variation of average deflection with incidence for the unmodified blades

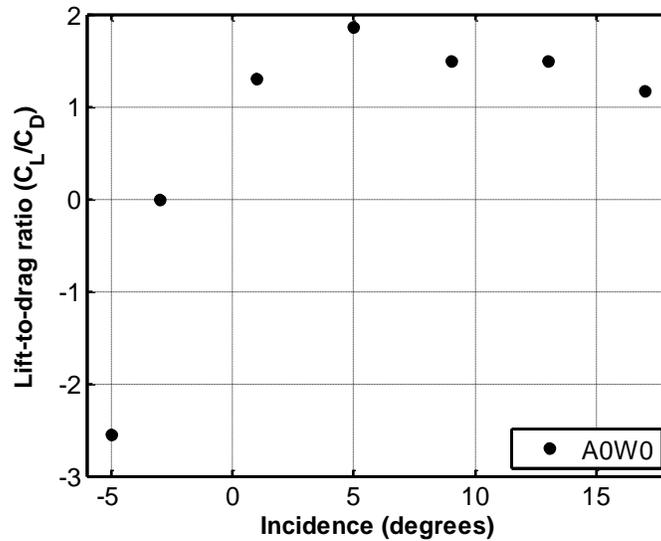

Figure 8: Variation of lift-to-drag ratio with incidence for the unmodified blades

### 3.2 Performance of cascade with modified blades

With the cascade with unmodified airfoil characterized, the set of blades were replaced with the different set of blades with tubercles as listed in Table 2. To directly compare the overall performance of a given blades set, the relative change in the performance parameters are plotted against incidence angles. Accordingly, a negative relative total pressure loss and a positive relative deflection is beneficial. Figure 9 shows the variation of the relative increase in the pitch-averaged total pressure loss coefficient for different incidence angles. Considering the uncertainty in the total pressure loss coefficient is about 0.03 (Table 3), the change from the baseline can be seen to be little or slightly positive for the low incidence angles. But for the incidence of 17°, a significant change can be seen for all the modified blades. At this incidence, for the blades A4W7.5 and A4W10 the tubercles appear to have resulted in total pressure recovery, whereas the blades A2W7.5 show a deterioration in terms of total pressure losses. All the blades show a slightly increased $\varsigma$ for the low incidence angles. With the exception of A2W7.5, all blades show a reversal of this general trend as the incidence is increased.

Figure 10 shows the relative increase in the average deflection for the different blade sets compared to the unmodified blades. Due to better certainty in the average deflection angle, a clear trend can be seen with change in both incidence and tubercles geometry. With changing incidence all blades show a sinuous trend, starting with positive or zero value from baseline, a decreasing trend, a negative minimum, an increasing trend, a positive maximum and finally a decreasing trend up to the highest incidence considered. This can be interpreted as follows: at zero incidence, the tubercles are affecting the flow adversely thus mostly reducing the deflection. As the incidence is increased, the tubercles affect the flow in a way to improve the deflection, although the effect appears to decline as incidence is further increased. Upon decreasing the incidence from zero, a similar effect improves the deflection. The camber of the blades is likely a contribution for the steep improvement from zero incidence in the negative incidence region compared to the positive incidence. The blades A4W10 show the best improvement both in the negative and positive high incidences, followed by A2W7.5 and A4W10 showing negligible improvement in terms of flow turning.

The variation of lift-to-drag ratio ($C_L/C_D$) with incidence is shown in Figure 11. The overall effect of tubercles is seen as a reduction in performance for low incidences and a small but finite improvement at higher, post-stall incidences. At the highest incidence considered, the percentage improvement for the blades A2W7.5, A4W7.5 and A4W10 are 0.9 %, 11.5 % and 17.9 % respectively. The performance of the blades A4W10 appears to be the best, evidently because of its improvement both in terms of total pressure recovery and flow turning at higher incidences as has been discussed earlier.

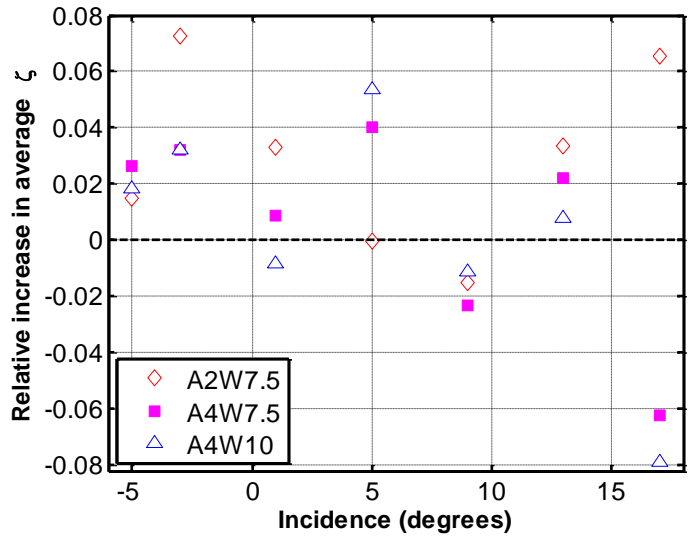

Figure 9: Comparison of relative total pressure loss coefficient with incidence

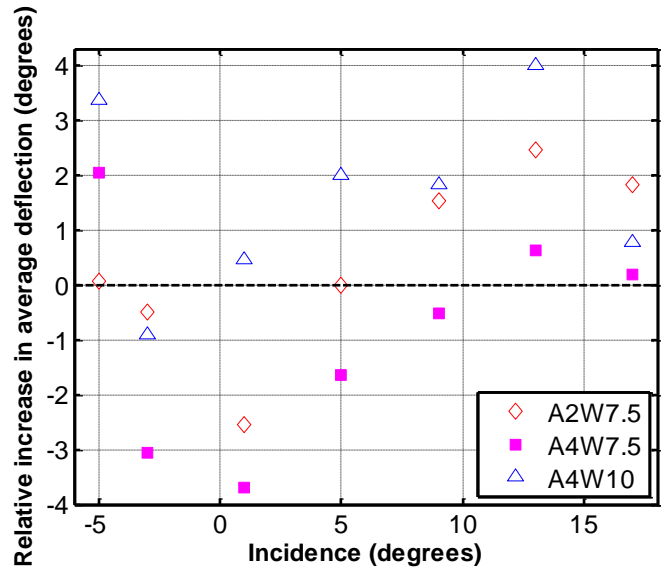

Figure 10: Comparison of relative average flow deflection with incidence

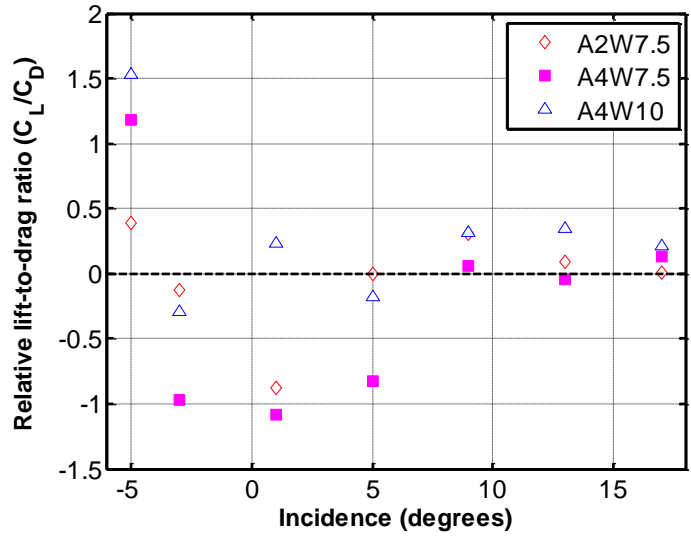

Figure 11: Variation of relative increase in lift-to-drag ratio with incidence

It is observed that the use of tubercles has improved the performance of the cascade for high incidence angles by improving total pressure recovery and flow deflection. For a two-dimensional compressor cascade analysis such as the one employed for this study, the total pressure loss is primarily due to the airfoil profile losses caused by boundary layer separation [1]. Therefore, it is likely that the presence of tubercles has changed the boundary layer separation characteristics. Figure 12 shows the estimated stall angles for each cascade. It can be seen that all the cascades with tubercles have resulted in delaying the stall. The cascade A2W7.5 shows the best stall improvement of up to 43%, despite showing poorer total pressure recovery. This suggests the importance of achieving complete turning to improve stall characteristics, since the extent of exit flow deviation is directly related to the losses.

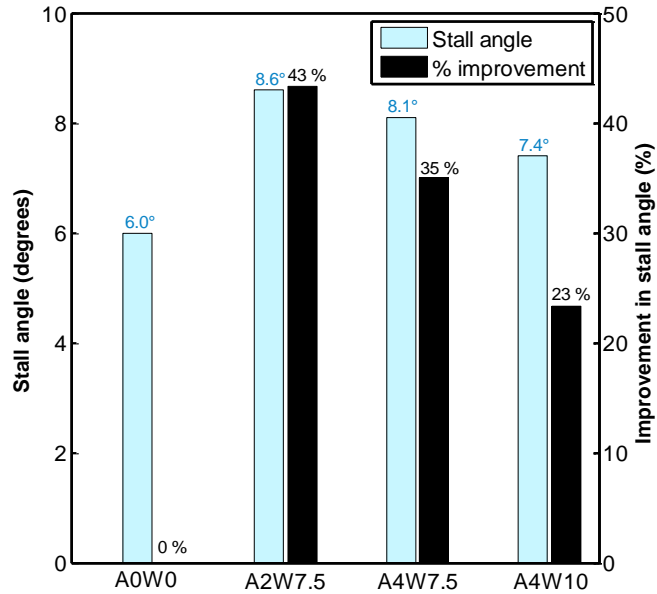

Figure 12: Estimated stall angle for different cascades

Some of the previously mentioned studies attribute this to one or more of the following effects: the formation of streamwise vortices that energize the boundary layer [10, 23], compartmentalization of flow in the span-wise direction [17, 25] and the lower pressure gradient caused due to the higher chord-length behind the tubercle crests [13, 19].

**3.2 Effect of tubercle geometry variation**

The effect of the amplitude of the tubercles can be understood by comparing the relative performance of A2W7.5 and A4W7.5. The total pressure loss coefficient indicates a lower loss for the blades with higher tubercle amplitude (Figure 9). However, from the flow deflection variation (Figure 10), we can see that the lower amplitude tubercles perform better due to a higher deflection. Due to the dependence of other performance parameters on both total pressure loss coefficient and deflection, and their opposing trend with respect to amplitude size, we can see contrasting performance in them. With respect to the total pressure loss, smaller amplitude tubercles seem to result in a better performance. This is consistent with the isolated airfoil performance in certain other studies [21, 23].

In order to evaluate the effect of wavelength, the cascade with tubercles A4W7.5 and A4W10 are compared. The total pressure loss coefficient plot (Figure 9) shows that the larger wavelength is slightly better. The average deflection also reflects that A4W10 is marginally better due to a higher deflection. The stall angle improvement is found to be better for the low wavelength tubercles case. It has to be noted that, for the tubercle dimensions in this study, the factor of decrease in tubercles wavelength size is 0.75, whereas for the amplitude it is 0.50. Similar non-sensitivity on wavelength has also been reported in previous studies [19, 21]. In the present study, the effectiveness of tubercles is found to be significantly dependent on amplitude and less dependent on the wavelength. This suggests that the difference in pressure gradient due to varying chord length perhaps plays a dominant role. This motivated high-resolution measurements in the span-wise direction whose results are discussed later.

The *A/W* ratio is a relevant parameter to examine the effect of tubercles acting as vortex generators, as the inclination angle is one of the main parameters of vortex generators. From Table 2, the tubercles in increasing order of *A/W* are A2W7.5, A4W10 and A4W7.5. There is no perceptible trend for *A/W* in the total pressure loss coefficient plot. From the flow deflection plot, the deflection appears to increase with decrease in *A/W* for the highest incidence, but a trend is absent at other incidences. The stall angle also shows no trend with respect to *A/W*. This suggests that compartmentalization of flow is more central than the vortex generator like effect of tubercles.

**3.3 Variation of flow properties across pitch in the exit plane**

The parameters considered so far, while useful in evaluating the performance of each blade at specified conditions, do not reveal the physics of the flow causing the changes. In order to infer the details of the blade-to-blade flow within the passage, the exit plane variations in the fundamental flow properties, namely, velocity components and total pressure are examined in this sub-section. To present the results in a concise manner without missing essential data, only a segment of the exit plane measurement plane will be shown here. The window of interest, as depicted in Figure 3(b), is a portion of the measurement plane and is located downstream of the central passage (bounded by blades 2 and 3), and is centered about the mid-span plane. The pitch-wise location of the window is maintained at a constant offset from the point of lowest measured total pressure in the wake of the second blade. Also, in order to appreciate the changes in magnitude the quantities are plotted in their dimensional form. Figure 13 shows the total pressure contours within the window of interest, arranged in an array spanning different tubercles along columns and for all incidence angles arranged row-wise. Figures 14 and 15 show similar plots for the dimensional axial and tangential velocities for all the tested cases.

The flow through the passage of the unmodified blades is considered first. From the total pressure contours, it can be seen that, up to an incidence of 5°, the total pressure increases especially near the pressure surface of the second blade. For higher incidences, the total pressure drops throughout the passage, with a greater drop near the suction surface of the third blade. With increasing incidence, the axial velocity contours show a trend of increase near the pressure surface and a decrease near the suction surface. The tangential velocity begins with a uniform distribution at slightly negative incidence and eventually, at the higher incidence angles, develops a gradient with a lower value near the pressure surface, and a less negative (closer to zero) near the mid passage and suction surface. All these three observations of the unmodified blades cascade are strongly suggestive of a detached flow on the suction surface at higher incidences, triggered after the incidence of 5°. The separation starts on the suction surface, and with increasing incidence, creates a partial blockage in the flow, reducing the axial velocity in that region, and to due to continuity, increasing it near the pressure surface. The presence of separation zone moves the fluid more towards the pressure surface, creating zones of negative tangential velocity as incidence increases.

Now, a comparison is made between the flow features of each of the modified blades with the baseline flow as created by the unmodified blades. Total pressure contours for the large-amplitude tubercles (A4W7.5 and A4W10) show larger separation zones at earlier incidences (9-13°), but have a reduced zone size at the highest incidence. The axial velocity also follows the total pressure trend, resulting in a more severe gradient in the pitch-wise direction at higher incidence. For all the cases, tubercles have made the tangential velocity lower near the pressure surface, at high incidence angles. For the tubercles A2W7.5, the total pressure loss is higher at the highest incidence, but due to better deflection, results in a better overall performance. It is to be noted that, for a given incidence, deflection angle depends on the axial and tangential velocity components. For blades A4W7.5 and A4W10, we see that the local span-averaged total pressure loss is increased in the wake regions, but is reduced in the region of the blade passage. But due to the larger area of the blade passage, the pitch-averaged total pressure loss is reduced.

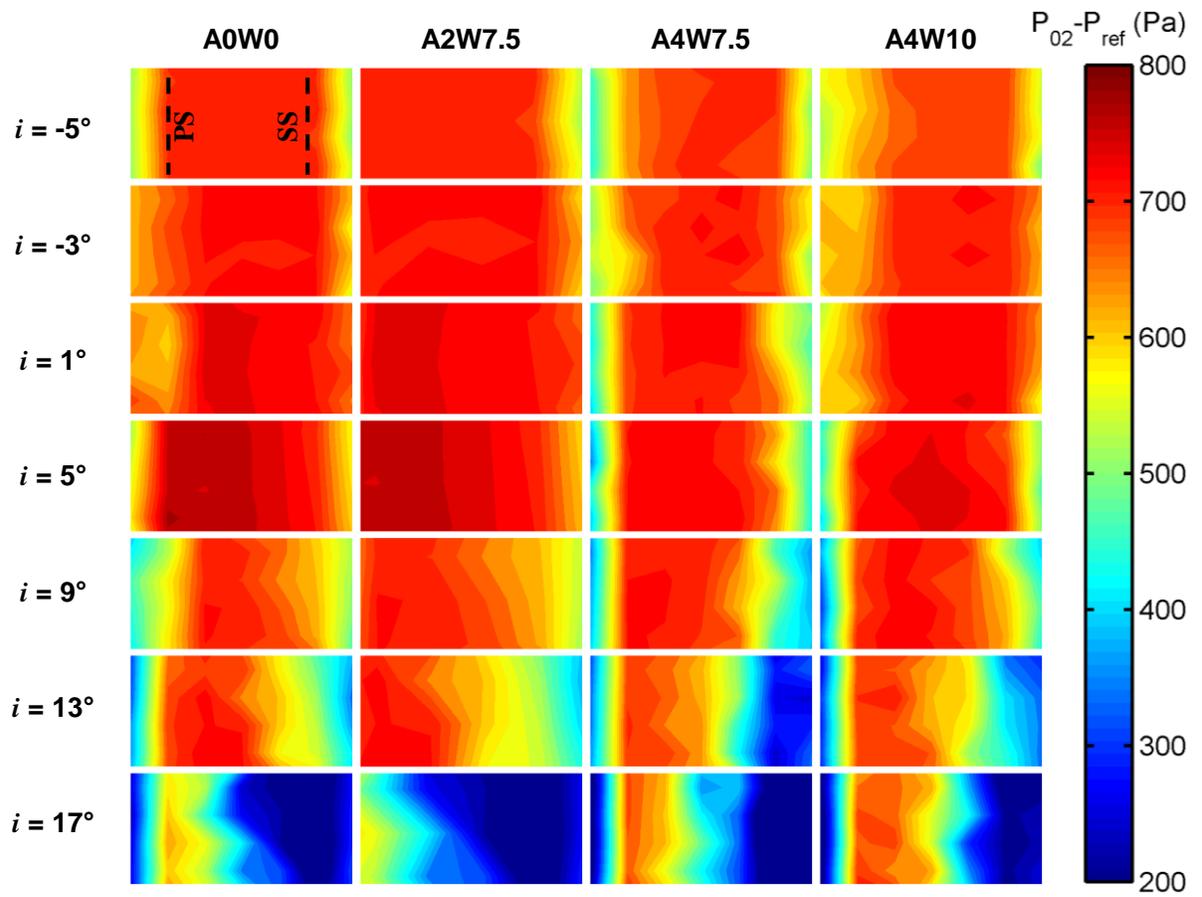

Figure 13: Total pressure contours for all blades and incidence angles, within the window of interest

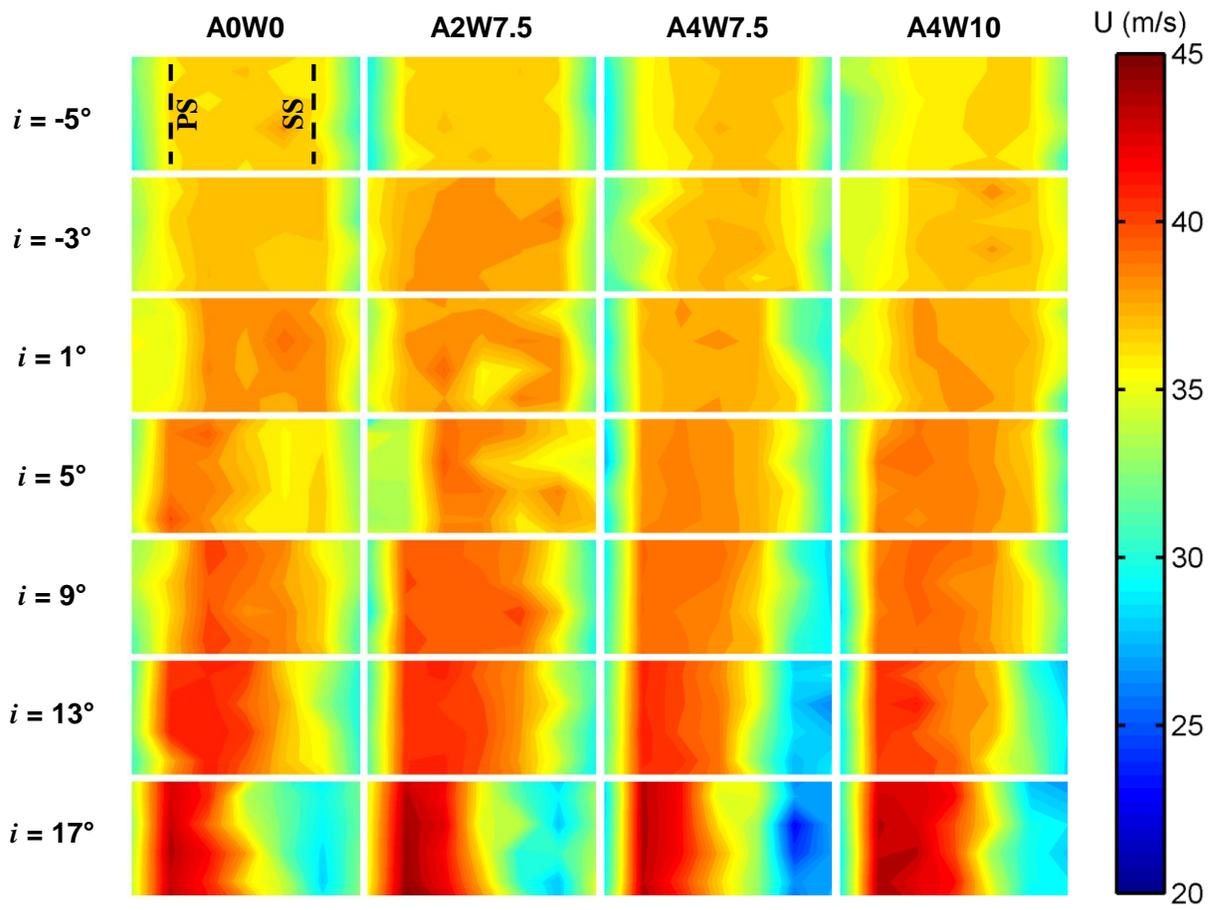

Figure 14: Axial velocity component contours for all blades and incidence angles, within the window of interest

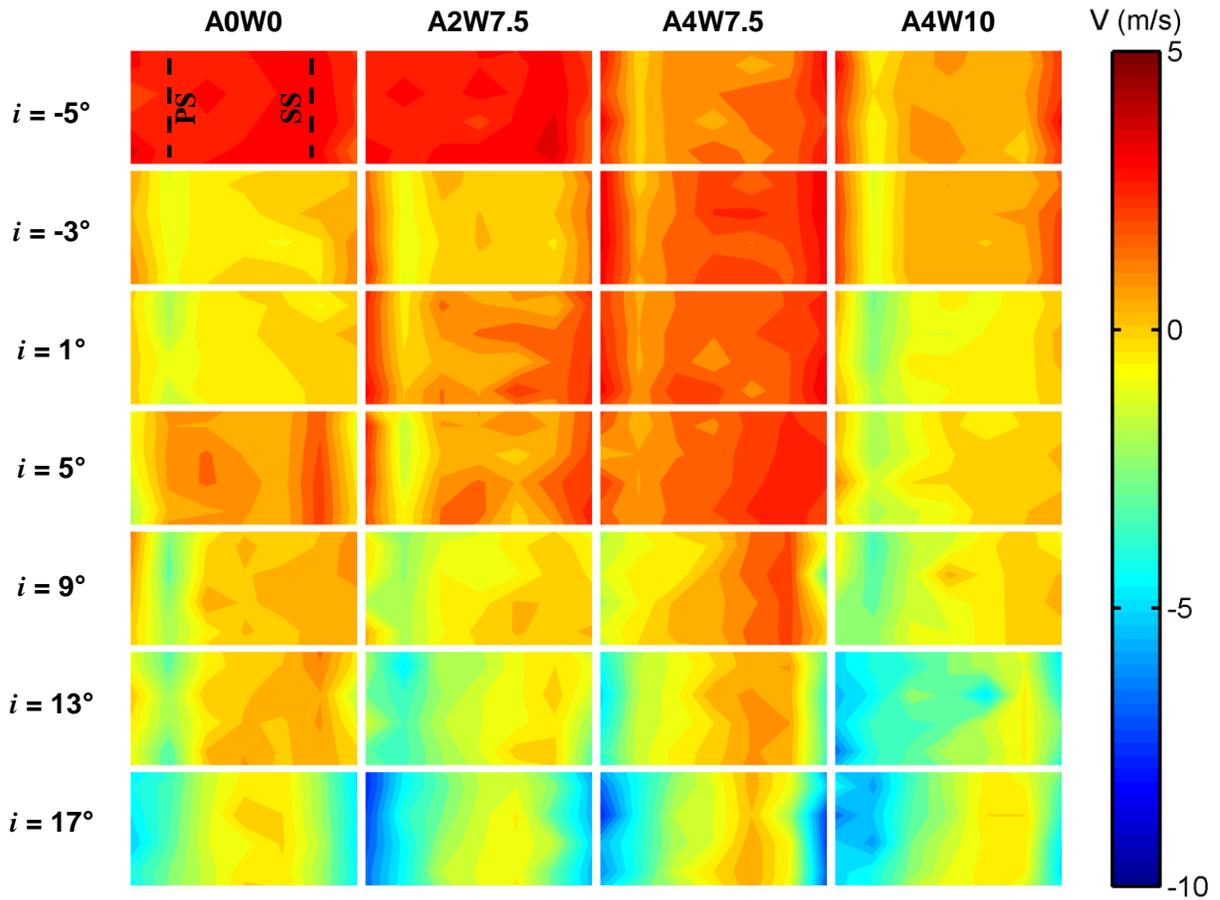

Figure 15: Tangential velocity component contours for all blades and incidence angles, within the window of interest

**3.4 Span-wise variation of velocity near blade surface**

In order to directly measure the flow profile in the blade-to-blade region, hot-wire measurements were carried out within the blade passage, along a line parallel to the span, just above the suction surface of the third blade, as shown in Figure 3(b). The probe was inserted from the upper end-wall of the cascade with the wire aligned so as to capture the axial velocity. The probe was located at a distance of quarter chord downstream of the leading edge, and the measurement line was about 2 mm from the suction surface of the third blade. The measurements were made approximately about the mid-span height. Figure 15 shows the plot of the variation of non-dimensional axial velocity (non-dimensionalized with the inlet freestream velocity) of all the blades with tubercles. All these measurements were made at an incidence of -3°.

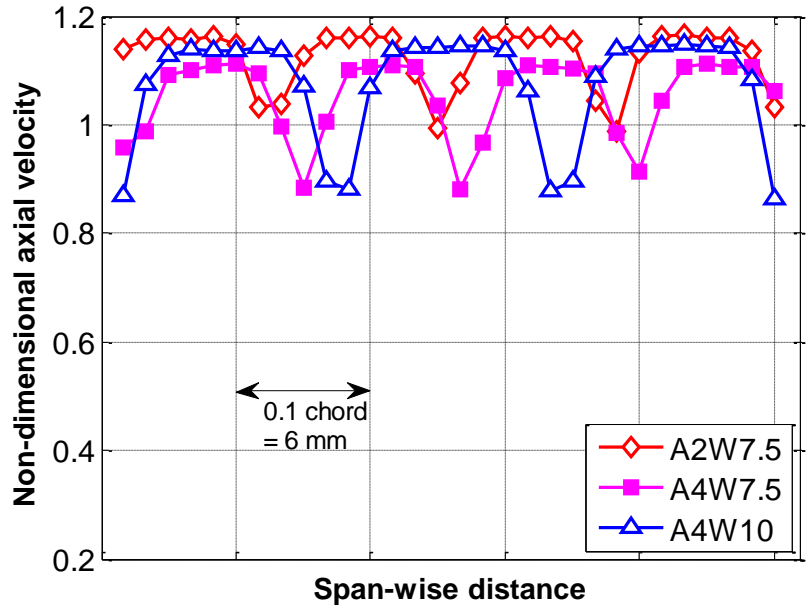

(a)

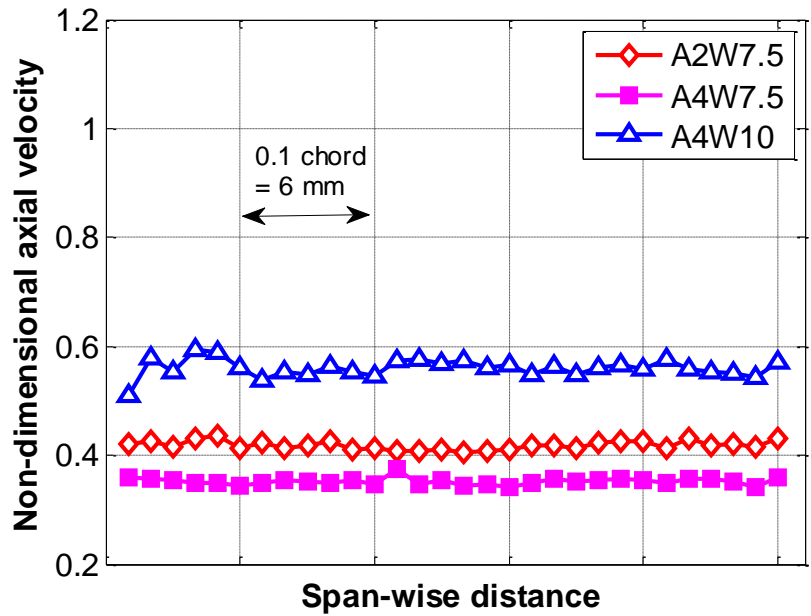

(b)

Figure 16: (a) Variation of axial velocity along the span, measured along a line located about 2 mm above the blade suction surface, quarter-chord from the leading edge

Figure 16 (a) shows a periodic variation of axial velocity along the span for all tubercles at quarter-chord from the leading edge. An inspection made at the time of measurements showed that the trough of the velocity profile (minimum velocity locations) were exactly downstream of the trough of the blade leading edge (smaller chord locations). A similar measurement 3 mm downstream of the trailing edge (Figure 16(b)) shows negligible peak-to-peak variation in the span-wise direction. Therefore, the effect of tubercles is to create a pair of mixing layer at each crest, in the plane perpendicular to the span-wise direction, in a region close to the blade surface. This is equivalent to the streamwise vortex sheet as identified by van Nierop et al. [19]. As the velocity gradients diminish just downstream of the blade, the five-hole probe measurement, which was carried out at half-cord downstream, cannot capture the gradients.

The peak-to-peak variation in axial velocity is seen to be dependent on the tubercle geometry: the smaller amplitude tubercles (A2W7.5) have a significantly smaller variation of about 0.15 compared to the larger ones (0.23 for A4W7.5 and 0.26 for A4W10). This is likely due to the difference in distances from the leading edge at the tubercle crest and trough to the quarter chord: the large pressure gradient on the suction surface results in measurably different velocities even for slightly different distances along the chord. Also, the peak velocity appears to follow a trend of increasing *A/W* ratio. As a result, a "sharper" tubercle seems to create a higher maximum velocity, but the peak-to-peak difference is higher for tubercles with "taller" tubercles. Since free shear layer properties depend on the ratio of the velocities of the two streams, the tubercle geometry can be expected to directly influence the performance of the cascade. We now see that the change in tubercle amplitude causes a far greater change in the ratio of velocity of the two streams, than the change in tubercle wavelength. This offers an explanation as to why the effect of amplitude is observed to be more substantial than that of the wavelength, for the geometries considered presently.

### 3.5 Flow Visualization

Surface oil-flow visualization has been carried out on the cascade end-wall for the cascade with the unmodified and modified blades. Only the central two blades are considered, as the flow around the outer blades are influenced by the side walls. It is observed that for low incidence angles, there is no perceptible difference in the oil streaks for the blades with and without tubercles. At higher incidence angles, a considerable variation is seen. Since only the qualitative flow description is of interest, the results presented here compare the unmodified blade with one representative blade with tubercles (A2W7.5), both set at an incidence of 17°. The other modified blades showed oil streak patterns qualitatively similar to that of A2W7.5. It is to be noted that the plane of the end walls coincided with the tubercle crests for all blades.

Figure 17 (a) shows the oil streak pattern for the unmodified blades at an incidence of 17°. The regions with accumulated pigment are indicative of separated flow. For the central two blades, it can be seen that flow separates on the suction side just downstream of the leading edge, marked by the arrows. Figure 17 (b) shows the oil streak pattern for the blades A2W7.5 for the same incidence. In this case, the streaklines can be seen to be attached for a greater distance along the chord, before lifting off the surface.

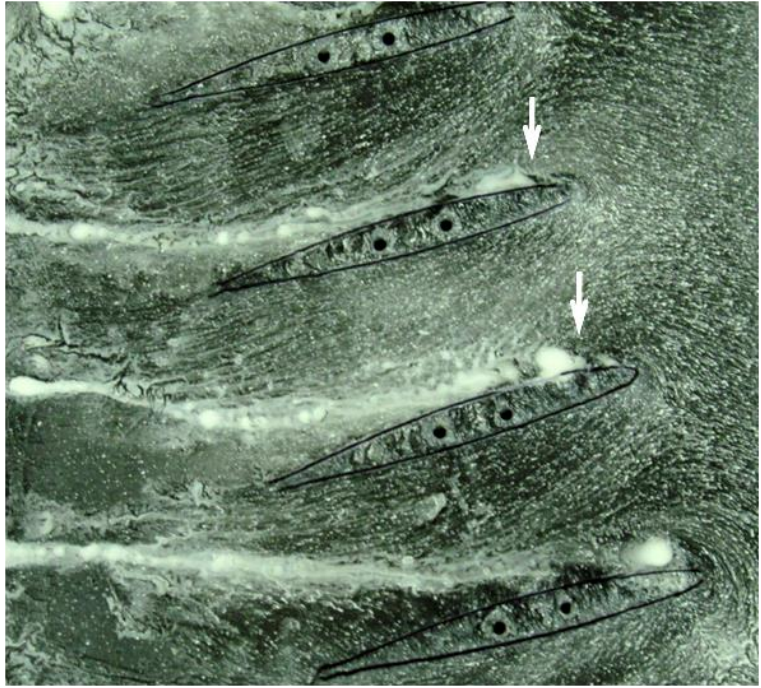

(a)

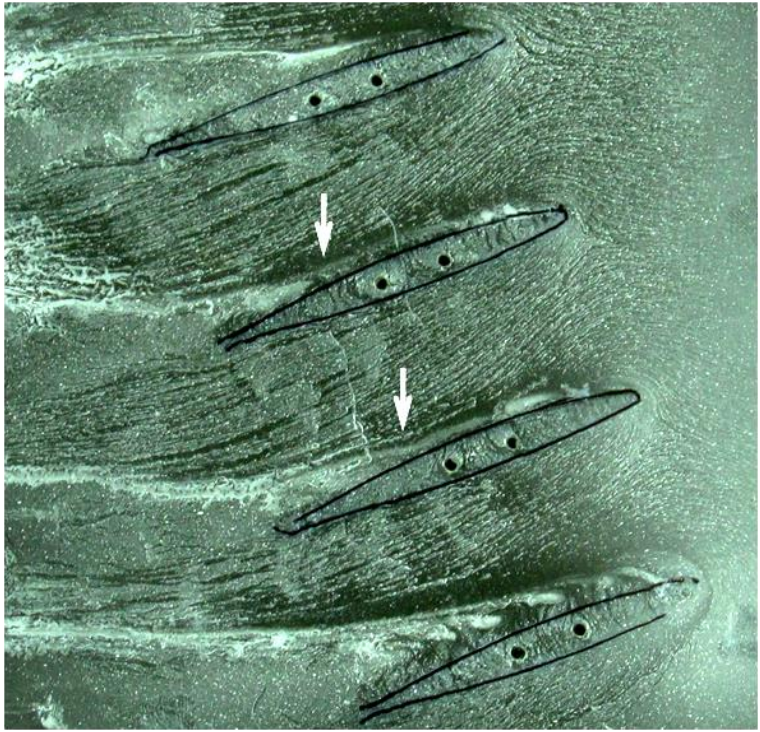

(b)

Figure 17: End wall oil streak pattern for the blades at $i = 17°$ for the cascade with (a) unmodified blades (b) blades having A2W7.5 tubercles. The arrows indicate the separation point on the suction surface of the middle two blades.

It must be noted that the end plane measurements cannot fully reveal blade-to-blade flow, although it helps in identifying the presence of certain features. This is because the flow significantly changes as we move downstream, in terms of thickening of the wakes, growth in boundary layer and mixing in the transverse directions. This can rationalize the apparent contradiction implied by the presence of a large separation region as shown by the total pressure contours downstream (Figure 13), and the small separation zone in the end-plane visualization (Figure 17(b)). Figures 14 and 15 also indicate that the flow angles are not highly angular downstream of the separation zone.

The results discussed thus far (Figures 6 through 17) effectively show that the mixing layers are caused due to the modified leading edge, resulting in significant mixing and momentum redistribution in the transverse directions. This helps in re-energizing the boundary layer along the suction surface of the blades, leading to reduced separated zone size and a redistribution of flow. In case of isolated airfoil applications, the increased loss due to total pressure is unfavorable, as it directly contributes to an increase in drag. For turbomachines, the performance in rotors is chiefly influenced by the pitch-averaged quantities. Also, for axial compressors, the delay of stall will enable considerably higher pressure recovery without the risk of entering into rotating stall. As a result, any means to increase the stall angle is extremely useful. This enables operation of blades at a high loading, which entails operation closer to the higher isentropic efficiency points on the compressor map.

Further investigations are needed to further examine the present findings with more detailed measurements. Some of the aspects which could not be addressed in this preliminary study include measurements over a finer grid and direct measurement of forces on the blade. The relatively large standard deviation exhibited at higher incidence indicates the importance of unsteady measurements in this study. PIV measurements over and downstream of the blades will facilitate in identifying the changes in the flow field caused due to tubercles. The benefit of tubercles seen in compressor blades also suggests possible advantages in turbine cascades, which are characterized by higher flow turning angles and possible secondary flows effects. Such a mechanism to reduce the effects of stalling or secondary flows may well result in reducing the number of stages in axial turbomachines.

## 4. CONCLUSIONS

The effect of sinusoidal leading edge tubercles in a linear compressor cascade with a Reynolds number of 130,000 is studied. A cascade with regular blades is first studied to establish the baseline performance and then blades with systematically varying tubercles geometries are tested. The primary performance parameters considered are the total pressure loss and the flow deflection angle. The results indicate that the performance of the cascade has substantially improved due to the effect of tubercles in terms of delaying the stall angle. The flow deflection is also found to be higher for all the blades with tubercles at the highest incidence angle. Based on these performance parameters, it is found that, at highest incidence angle considered (17°), the overall improvement is the highest for the blades with smaller tubercles amplitude and wavelength. For moderately high incidences, the tubercles with larger amplitudes gave a better performance. The surface oil flow visualization on the end-wall reveals a delay in the onset of flow separation. The dominant cause of separation delay is likely due to the reduction in flow separation tendency caused by the reduced pressure gradient behind the tubercle crests. The resulting mixing that is associated with mixing layers of different velocities can be considered to be the cause of a series of effects leading to change in boundary layer separation characteristics and eventually a measurable change in the performance of the cascade. A blade with such substantially improved stall angles will enable axial compressors to operate with improved pressure ratios without the risk of blade stall.